\documentclass[12pt]{article}
\usepackage{epsf,amsfonts,latexsym}
\newsymbol\ltimes 226E
\newsymbol\rtimes 226F
\epsfverbosetrue
\def\double{\mathbb}

\def\cc{{\double C}}
\def\nn{{\double N}}
\def\rr{{\double R}}
\def\zz{{\double Z}}

\def\aa{{\cal A}}
\def\ccc{{\cal C}}

\def\hh{{\cal H}}

\def\aa{{\cal A}}
 
\def\hh{{\cal H}}

\def\lll{{\cal L}}

\def\t{{\rm tr}\,}

\def\ddd{{\,\hbox{$\partial\!\!\!/$}}}

\def\de{\hbox{\rm d}}

\def\pa{{\partial}}

\def\lb{\left[} 
\def\rb{\right]}

\def\op{\oplus}

\def\bb{\begin{eqnarray}}
\def\ee{\end{eqnarray}}
\def\eee{\nonumber\end{eqnarray}}
\def\pp{\pmatrix}
\def\qq{\quad}

\begin{document}

\hsize 17truecm
\vsize 24truecm
\font\twelve=cmbx10 at 18pt
\font\eightrm=cmr8
\baselineskip 18pt

\noindent
{\twelve Noncommutative geometry and the standard model}
{ Encyclopedia of Mathematical
Physics}, J.-P. Franoise, G. Naber \& Tsou Sheung
Tsun (eds.), Elsevier Science
\section{Introduction}

The aim of this contribution is to explain how Connes derives the
standard model of electromagnetic, weak and strong forces  from
noncommutative geometry. The reader is supposed to be
aware of two other derivations in fundamental physics: the
derivation of the Balmer-Rydberg formula for the spectrum of the
hydrogen atom from quantum mechanics and Einstein's derivation of
gravity from Riemannian geometry.

At the end of the 19th century, new physics was discovered in atoms,
their discrete spectra. Balmer and Rydberg succeeded to put order into
the fast growing set of experimental numbers with the help of a
phenomenological ansatz for the frequencies $\nu $ of the spectral
rays of e.g. the hydrogen atom,
\bb \nu =g(n_2^q-n_1^q),\qq n_j\in\nn,\qq q\in\zz,\qq g\in\rr.\ee
The integer variables $n_1$ and $n_2$ reflect the
discreteness of the spectrum. On the other hand the discrete parameter
$q$ and the continuous parameter $g$ were fitted by experiment:
$q=-2$ and $g=3.289\ 10^{15}$ Hz, the famous Rydberg constant.
Later quantum mechanics was discovered and allowed to derive the
Balmer-Rydberg ansatz and to constrain its parameters:
\bb q=2\qq {\rm and} \qq g=\,\frac{m_e}{4\pi \hbar^3}\,
\frac{e^4}{(4\pi \epsilon_0)^2}\, ,\ee
in beautiful agreement with the anterior experimental fit.

\section{The standard model}

We propose to introduce the standard model in analogy with the
Balmer-Rydberg formula, Tab. 1.

\subsection{The Yang-Mills-Higgs ansatz}

The variables of this Lagrangian ansatz are spin 1 particles $A$, spin
${\textstyle\frac{1}{2}} $ particles
decomposed into left- and right-handed components $\psi =(\psi
_L,\psi _R)$ and spin 0 particles $\varphi $. There are four discrete
parameters, a compact real Lie group $G$, and three unitary
representations on complex Hilbert spaces $\hh_L,\ \hh_R$, and
$\hh_S$. The spin 1 particles come in a multiplet living in the
complexified  of the Lie algebra of $G$, $A\in{\rm
Lie}(G)^\cc$. The left-handed and right-handed spinors
come in multiplets living in the Hilbert spaces,
$\psi _L\in\hh_L,\ \psi _R\in\hh_R$. The (Higgs) scalar is another
multiplet, $\varphi \in\hh_S$. The Yang-Mills-Higgs Lagrangian
together with its Feynman
diagrams is spelled out in Tab. 2.

There are several continuous parameters:
the gauge coupling $g\in\rr_+$,
the Higgs self-couplings $\lambda ,\mu \in\rr_+$ and a bunch of
Yukawa couplings $g_Y\in\cc$.

Let us choose $G=U(1)\owns e^{i\theta }$. Its irreducible unitary
representations are all 1-dimensional, $\hh=\cc\owns \psi $
characterized by the charge $q\in\zz$: $\rho (e^{i\theta })\psi =
e^{iq\theta }\psi$. Then with $q_L=q_R$ and $\hh_S=\{0\} $ we get
Maxwell's theory with the photon (or gauge boson or 4-potential)
$A$ coupled to the Dirac theory of a massless spinor  of electric
charge
$q_L$  whose (relativistic) wave function is $\psi $. The gauge
coupling is given by
$g=e/\sqrt{\epsilon_0}.$ Gauge invariance of the
Yang-Mills-Higgs Lagrangian implies via Emmy Noether's
theorem electric charge conservation in this case.

Yang-Mills models are therefore simply nonAbelian generalizations of
electromagnetism where the Abelian gauge group $U(1)$ is replaced
by any compact real Lie group. We insist on compact because all
irreducible unitary representations of compact groups are finite
dimensional. Finally the Higgs scalar is added to give masses to spinors
and gauge bosons via spontaneous symmetry breaking.

We use compact groups and unitary representations as
(discrete) parameters. One motivation is Noether's theorem and
conserved quantities. The other comes from Wigner's theorem: The
irreducible unitary representations of the Poincar\'e group are
classified by mass and spin. Its orthonormal basis vectors are
classified by energy-momentum and by the $z$-component of
angular momentum. This theorem leads to the widely accepted
definition of a particle as an orthonormal basis vector in a Hilbert
space
$\hh$ carrying a unitary representation $\rho $ of a group $G$.

A precious property of the Yang-Mills-Higgs ansatz is its
perturbative renormalizability  necessary for fine
structure calculations like the anomalous magnetic moment of the
muon.

\subsection{The experimental fit}

Physicists have spent some thirty years and some $10^9$
Swiss Francs to distill the fit \cite{data}:
\bb G&=&SU(2)\times U(1)\times
SU(3)/(\zz_2\times\zz_3),\label{smgr}\\[2mm]
\hh_L &=& \bigoplus_1^3\lb
(2,{\textstyle\frac{1}{6}},3)\op
(2,-{\textstyle\frac{1}{2}},1)
\rb  ,\label{hl}\\
\hh_R& = &\bigoplus_1^3\lb
(1,{\textstyle\frac{2}{3}},3)\oplus
(1,-{\textstyle\frac{1}{3}},3)\op (1,-1,1)
\rb,\label{hr} \\
\hh_S &= &(2,-{\textstyle\frac{1}{2}},1)\label{hs}.
\ee
Here $(n_2, y, n_3)$
denotes the tensor product of an $n_2$ dimensional
representation of $SU(2)$, `(weak) isospin', an $n_3$ dimensional
representation of $SU(3)$, `colour', and the one dimensional
representation of $U(1)$ with `hyper'charge $y$.  For
historical reasons the hypercharge is an integer
multiple of ${\textstyle\frac{1}{6}}$.
This is irrelevant: in the Abelian case, only the product of the
hypercharge with its gauge coupling is measurable and we do not
need multi-valued representations, which are characterized by
non-integer, rational hypercharges.
In the direct sum, we recognize the three
generations of fermions, the quarks, `up, down, charm, strange, top,
bottom', are
$SU(3)$
 triplets, the leptons, `electron, $\mu $, $\tau $' and their neutrinos,
are colour singlets. The basis of the fermion representation space is
\bb \pp{u\cr d}_L,\ \pp{c\cr s}_L,\ \pp{t\cr b}_L,\
\pp{\nu_e\cr e}_L,\ \pp{\nu_\mu\cr\mu}_L,\
\pp{\nu_\tau\cr\tau}_L\eee
\bb\matrix{u_R,\cr d_R,}\qq \matrix{c_R,\cr s_R,}\qq
\matrix{t_R,\cr b_R,}\qq  e_R,\qq \mu_R,\qq
\tau_R\eee
The
parentheses indicate isospin doublets.

The eight gauge bosons associated to
$su(3)$ are called gluons. Attention, the $U(1)$ is not the one of electric
charge, it is called hypercharge, the electric charge
is a linear combination of hypercharge and weak
isospin. This mixing is
necessary to give electric charges to the $W$ bosons.
The $W^+$ and $W^-$ are pure isospin states, while the
$Z^0$ and the photon are (orthogonal) mixtures of the
third isospin generator and hypercharge.

As the group $G$ contains three simple factors there are three gauge
couplings, 
\bb g_2=0.6518\pm 0.0003,&g_1=0.3574\pm 0.0001,&
g_3=1.218\pm 0.01,\label{gaugecoup}\ee
The Higgs couplings are usually expressed in terms of of the $W$- and
Higgs-masses:
\bb m_W&=&{\textstyle\frac{1}{2}}g_2\,v
\,=\,80.419\pm 0.056\ {\rm GeV},\\
m_\varphi &=&2\sqrt 2\sqrt\lambda\,v\,>\,98\ {\rm GeV},\ee
with the vacuum expectation value
$v:={\textstyle\frac{1}{2}}\mu/\sqrt\lambda$.
Because of the high
degree of reducibility of the spin ${\textstyle\frac{1}{2}} $
representations there are  27 complex Yukawa
couplings. They constitute the fermionic mass matrix which contains
the fermion masses and mixings.
\bb m_e=0.510998902\pm 0.000000021\ {\rm MeV},&
m_u=3\pm 2\ {\rm MeV},&m_d=6\pm 3\ {\rm MeV},
\cr
m_\mu=0.105658357\pm 0.000000005\ {\rm GeV},&
m_c=1.25\pm 0.1\ {\rm GeV},&
m_s=0.125\pm 0.05\ {\rm GeV},\cr
m_\tau=1.77703 \pm 0.00003\ {\rm GeV},&
m_t=174.3\pm 5.1\ {\rm GeV},&
m_b=4.2\pm 0.2\ {\rm GeV}.\eee
For simplicity, we have taken massless neutrinos. Then  mixing only
occurs for quarks and is given by
a unitary matrix, the Cabibbo-Kobayashi-Maskawa
matrix
\bb C_{KM}:=\pp{V_{ud}&V_{us}&V_{ub}\cr
V_{cd}&V_{cs}&V_{cb}\cr  V_{td}&V_{ts}&V_{tb}},\ee
whose matrix elements in absolute value are:
\bb \pp{
0.9750\pm 0.0008&0.223\pm 0.004&0.004\pm
0.002\cr
0.222\pm 0.003&0.9742\pm 0.0008&0.040\pm 0.003\cr
0.009\pm 0.005&0.039\pm 0.004&0.9992\pm 0.0003}.
\ee
The physical meaning of the quark mixings is the
following: when a sufficiently energetic $W^+$ decays
into a $u$ quark, this
$u$ quark is produced together with a
$\bar d$ quark with probability $|V_{ud}|^2$, together
with a
$\bar s$ quark with probability $|V_{us}|^2$, together
with a
$\bar b$ quark with probability $|V_{ub}|^2$. 

The phenomenological success of the standard model
is phenomenal: with only a handful of parameters, it reproduces
correctly some millions of experimental numbers: cross sections, life
times, branching ratios.

\section{Noncommutative geometry}

Noncommutative geometry is an analytical geometry generalizing
three other ones that also had important impact on our understanding
of forces and time. Let us start by briefly recalling the three 
forerunners, Tab. 3. {\bf
Euclidean geometry} is underlying Newton's mechanics as space of
positions. Forces are described by vectors living in the same space
and the Euclidean scalar product is needed to define work and
potential energy. Time is not part of geometry, it is absolute. This
point of view is abandoned in special relativity unifying space
and time into {\bf Minkowskian geometry}. This new point of view
allows to derive the magnetic field from the electric field as a
pseudo force associated to a Lorentz boost. Although time has
become relative, one can still imagine a grid of synchronized
clocks, i.e. a universal time. The next generalization is
{\bf Riemannian geometry} = curved spacetime. Here gravity can be
viewed as the pseudo force associated to a uniformly accelerated
coordinate transformation. At the same time, universal time loses
all meaning and we must content ourselves with proper time. With
today's precision in time measurement, this complication of life
becomes a bare necessity, e.g. the global positioning system (GPS).

 Our last
generalization is   noncommutative geometry = curved
space(time) with an uncertainty principle.  As in quantum mechanics,
this uncertainty principle is introduced via noncommutativity.

\subsection{Quantum mechanics}

Consider the classical harmonic oscillator. Its phase space is
$\rr^2$ with points labelled by position $x$ and momentum $p$. A
classical observable is a differentiable function on phase
space such as the total energy $p^2/(2m)\,+\,kx^2$.
Observables can be added and multiplied, they form the algebra
$\ccc^\infty(\rr^2)$, which is associative and commutative. To pass
to quantum mechanics, this algebra is rendered noncommutative
by means of a noncommutation relation for the
generators
$x$ and $p$:
$ [x,p]=i\hbar 1.$
Let us call $\aa$ the resulting algebra `of quantum observables'. It
is still associative, has an involution $\cdot^*$ (the adjoint or Hermitian
conjugation)  and a unit 1.

Of course, there is no space anymore of which
$\aa$ is the algebra of functions. Nevertheless, we talk about such a
`quantum phase space' as a space that has no points or a space with
an uncertainty relation. Indeed, the noncommutation relation implies
Heisenberg's uncertainty relation
$ \Delta x\Delta p \geq \hbar /2$
and tells us that points in phase space lose all meaning, we can
only resolve cells in phase space of volume $\hbar/2$, see Fig. 1.
To define the uncertainty $\Delta a$ for an observable $a\in\aa$,
we need a faithful representation of the algebra on a Hilbert
space,  i.e. an injective homomorphism
$\rho $ from $\aa$ into the algebra of operators on $\hh$.
For the harmonic oscillator, this Hilbert space is
$\hh=\lll^2(\rr)$. Its elements are the wave functions $\psi (x)$,
square integrable functions on configuration space. Finally,
the dynamics is defined by the Hamiltonian, a self adjoint observable
$H=H^*\in\aa$ via Schr\"odinger's equation
$ \left( i\hbar \pa / \pa t -\rho (H)\right) \psi
(t,x)=0.$
Here time is
 an external parameter, in particular, time is not
 an observable. This is different in the special
relativistic setting where Schr\"odinger's equation is replaced by
Dirac's equation
$\ddd\psi =0.$
Now the wave function $\psi $ is the four-component spinor
consisting of left- and right-handed, particle and antiparticle
wave functions.  The Dirac operator is not in
$\aa$ anymore, it is still an operator on $\hh$. In Euclidean
spacetime, the Dirac operator is also self adjoint,
$\ddd^*=\ddd.$

\subsection{Spectral triples}

Noncommutative geometry \cite{connes} does to a compact Riemannian
spin manifold  $M$ what quantum mechanics does to phase space.
A noncommutative geometry is defined by the three purely algebraic
items, $(\aa,\hh,\ddd)$ called a spectral triple.  $\aa$ is a real,
associative, possibly noncommutative involution algebra with unit,
faithfully represented on a complex Hilbert space $\hh$, and $\ddd$ is a
self adjoint operator on $\hh$.
As the spectral triple, also the axioms linking its three items  are
motivated by relativistic quantum mechanics. Connes reconstruction
theorem (1996, 
\cite{grav}) states that there is a one-to-one correspondence between
spectral triples with {\it commutative} algebra $\aa$ and Riemannian
manifolds $M$. As for classical phase space, the algebra consists of
differentiable functions now on the Riemannian manifold,
$\aa=\ccc^\infty(M)$. The algebra is represented on spinors on which
the Dirac operator acts. As for  quantum phase space, Connes
defines a noncommutative geometry by a spectral triple whose algebra
is allowed to be noncommutative and he shows how important properties
like dimensions, direct products, differentiation, integration and
distances generalize to the noncommutative setting. As a bonus, the
algebraic axioms of a spectral triple, commutative or not, include
discrete i.e. 0-dimensional spaces that now are naturally equipped with
a differential calculus. These spaces have finite dimensional algebras
and Hilbert spaces meaning that their algebras are just matrix
algebras. 

\subsection{The spectral action}

In a next step Connes \& Chamseddine \cite{grav,cc} consider
noncommutative spacetimes.
They define the spectral action, a
generalization of the Einstein-Hilbert action to noncommutative
spacetimes, and compute it explicitly for almost commutative
geometries.

An almost
commutative geometry is defined as  a direct product of a
$4$-dimensional commutative geometry, `ordinary spacetime', by a
0-dimensional noncommutative geometry, the `internal space'. If the
latter is also commutative, e.g. the ordinary two point space, then the
direct product describes a two-sheeted universe or a Kaluza-Klein space
whose fifth dimension is discrete, \cite{john}. In general, the axioms
of spectral triples imply that the `Dirac' operator of the internal space is
precisely the fermionic mass matrix.

On almost commutative geometries,
the spectral action is equal to the Einstein-Hilbert action plus the
Yang-Mills-Higgs ansatz, Fig. 2. In other words, noncommutative
geometry explains the forces mediated by gauge bosons and Higgs
scalars as pseudo-forces accompanying the gravitational force in the
same way that Minkowskian geometry (i.e. special relativity) explains
the magnetic force as a pseudo-force accompanying the electric force.

There are constraints on the discrete and continuous parameters in the
Yang-Mills-Higgs ansatz deriving from the spectral action Fig. 3.

In particular if we consider only irreducible spectral triples and among
them only those which produce non-degenerate fermion masses
compatible with renormalization then we only get the standard model
with one generation of quarks and leptons, with a massless neutrino
and with an arbitrary number of colours, and a few submodels thereof.
More than one generation and neutrino masses are possible but imply
reducible triples. However in at least one generation,
the neutrino must remain purely left and massless.

For the standard model with $N$ generations and $N_c$
colours, we have the constraints
$g_{N_c}^2=g_2^2=(9/N) \lambda$ on the continuous parameters. If we
put
$N=N_c=3$ and if we believe in the popular `big desert' then these
constraints yield a `unification scale'  $\Lambda = 10^{17}$ GeV at
which the uncertainty relation in spacetime should become manifest,
$\Delta \tau =\hbar /\Lambda $, and a Higgs-mass of
$m_\varphi =171.6
\pm 5$ GeV for $m_t= 174.3\pm 5.1$ GeV, see Fig. 4.

It is clear that almost commutative geometries only scratch the surface
of a gold mine. May we hope that a genuinely
noncommutative geometry will solve our present problems with
quantum field theory and quantum gravity??

\noindent
Thomas Sch\"ucker\\
Centre de Physique Th\'eorique\\
case 907, 
F-13288 Marseille, cedex 9\\
also at Universit\'e de Provence\\
schucker@cpt.univ-mrs.fr

\vskip1cm

\begin{table}[h]
\begin{center}
\begin{tabular}{l|ll}
&${}$\qq atomic physics& particle physics\\[1ex]
\hline &\\
new physics&${}$\qq discrete spectra&forces mediated by gauge
bosons\\[1ex]
ansatz &${}$\qq  $\nu =g(n_2^q-n_1^q)$& Yang-Mills-Higgs
models \\[1ex] 
experimental fit&${}$\qq $q=-2,\ g=3.289\ 10^{15}$ Hz\qq ${}$
&standard model\\[1ex]
underlying theory \qq ${}$ &${}$\qq  quantum
mechanics&noncommutative geometry
\end{tabular}
\caption{An analogy}
\end{center}
\end{table}

\begin{table}[h]
\begin{center}
\begin{tabular}{clr}
$\lll[A,\psi ,\varphi ]=$&${}\ {\textstyle\frac{1}{2}}\, \t(\pa_\mu
A_\nu
\pa^\mu A^\nu -
\pa_\mu A_\nu \pa^\nu A^\mu)$& \epsfxsize=1cm
\epsfysize=1cm
\epsfbox{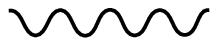}\\[1ex]
&$+g\ \t (\pa_\mu A_\nu [A^\mu ,A^\nu ])$&\epsfxsize=1cm
\epsfysize=1cm
\epsfbox{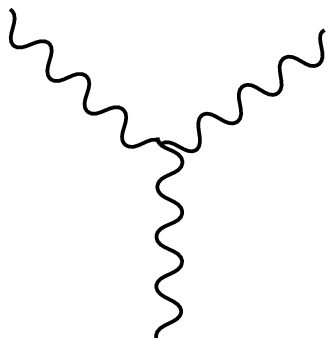}\\[1ex]
&$+g^2\,\t ([A_\mu ,A_\nu ][A^\mu ,A^\nu ])$&\epsfxsize=1cm
\epsfysize=1cm
\epsfbox{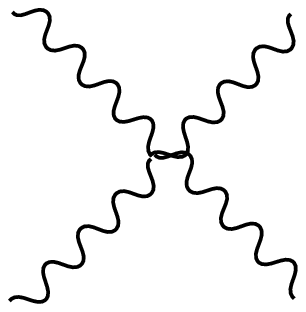}\\[1ex]
&&\\[1ex]
&$ +\bar \psi \ddd\psi  $&\epsfxsize=1cm
\epsfysize=1cm
\epsfbox{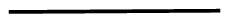}\\[1ex]
&$ +i g\,\bar\psi (\tilde\rho _L\op\tilde\rho _R)(A_\mu )\,\gamma
^\mu \psi  $&\epsfxsize=1cm
\epsfysize=1cm
\epsfbox{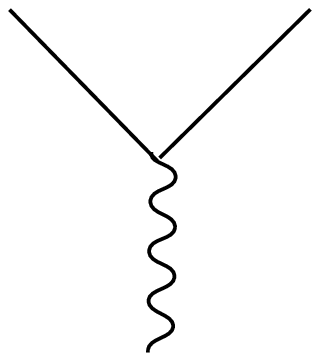}\\[1ex]
&&\\[1ex]
&$+ {\textstyle\frac{1}{2}} \,\pa_\mu \varphi ^*\pa^\mu\varphi  $
&\epsfxsize=1cm
\epsfysize=1cm
\epsfbox{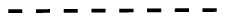}\\[1ex]
&$+ {\textstyle\frac{1}{2}} \,g\,\{(\tilde \rho _S(A_\mu )\varphi)
^*\pa^\mu \varphi  + \pa_\mu \varphi ^*
\tilde \rho _S(A_\mu )\varphi \}$ &\epsfxsize=1cm
\epsfysize=1cm
\epsfbox{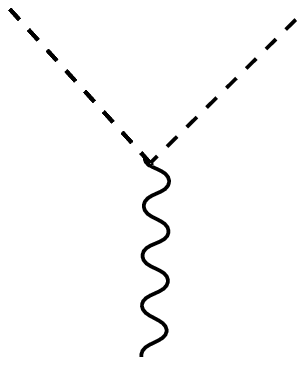}\\[1ex]
&$ + {\textstyle\frac{1}{2}} \,g^2\,(\tilde \rho _S(A_\mu )\varphi)
^*\tilde \rho _S(A^\mu )\varphi$
&\epsfxsize=1cm
\epsfysize=1cm
\epsfbox{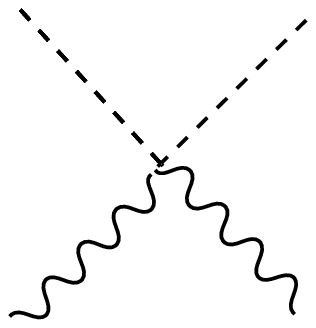}\\[1ex]
&$ +\lambda \, \varphi ^*\varphi \varphi ^*\varphi
$&\epsfxsize=1cm
\epsfysize=1cm
\epsfbox{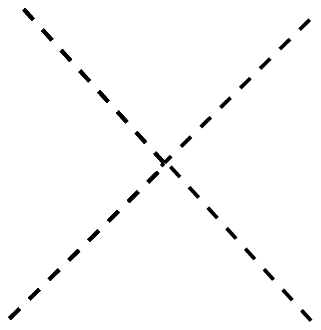}\\[1ex]
&&\\[1ex]
&$ -{\textstyle\frac{1}{2}} \,\mu ^2\,\varphi ^*\varphi
$&\epsfxsize=1cm
\epsfysize=1cm
\epsfbox{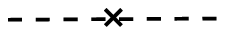}\\[1ex]
&$ +\,g_Y\,\bar\psi \varphi \psi
+\,\bar g_Y\,\bar\psi \varphi ^* \psi $&\epsfxsize=1cm
\epsfysize=1cm
\epsfbox{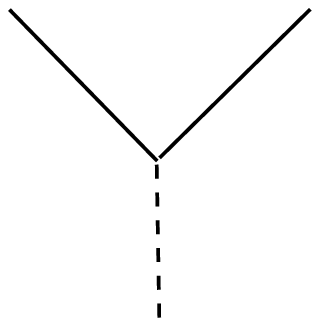}
\end{tabular}
\caption{The Yang-Mills-Higgs Lagrangian and its Feynman
diagrams}
\end{center}
\end{table}

\begin{table}[h]
\begin{center}
\begin{tabular}{l|ll}
geometry&force & time\\[1ex]\hline &\\
Euclidean &$E=\int\vec F\cdot\de \vec x$&absolute\\[1ex]
Minkowskian&$\vec E,\epsilon _0\Rightarrow\vec B,\mu
_0=\,\frac{1}{\epsilon _0 c^2}\, $&universal\\[1ex]
Riemannian&Coriolis $\leftrightarrow$ gravity&proper, $\tau$
\\[1ex]
noncommutative\qq\qq&gravity $\Rightarrow$ YMH, $\lambda
={\textstyle\frac{1}{3}} g_2^2$\qq\qq&$\Delta \tau \sim 10^{-40}$
 s
\end{tabular}
\end{center}
\caption{Four nested analytic geometries}
\end{table}

\begin{figure}[h]
\hspace{5.5cm}
\setlength{\unitlength}{1.0cm}
\begin{picture}(10,6.5)(0.5,0)
\put(3,2){\framebox(1,1)}
\put(0,0.5){\vector(1,0){6}}
\put(0.5,5.3){\parbox[b]{2cm}{$p$}}
\put(0.5,0){\vector(0,1){5}}
\put(6.2,0.4){\parbox{1cm}{$x$}}
\put(4.15,2.7){\parbox{1cm}{$\hbar/2$}}
\put(3.4,2.6){\circle*{0.1}}
\end{picture}
\caption{The first example of noncommutative geometry}
\end{figure}
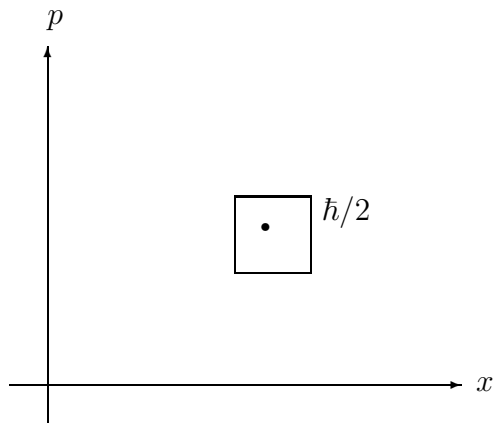

\begin{figure}[h]
\begin{center}
\setlength{\unitlength}{1.0cm}
\begin{picture}(10,8)(0,1)
\put(0,1){\parbox{2cm}{\rm Riemannian geometry}}
\put(4,1){\vector(1,0){2,5}}
\put(4.5,1.5){\parbox[b]{2cm}{\rm Einstein}}
\put(7.5,1){\parbox{2cm}{\rm gravity}}

\put(1.4,2.1){\oval(0.2,0.6)[b]}
\put(1.3,2.1){\vector(0,1){.5}}
\put(1.2,3.5){\parbox{2.5cm}{\rm almost\\ commutative
geometry}}
\put(1.4,4.9){\oval(0.2,0.6)[b]}
\put(1.3,4.9){\vector(0,1){2.2}}

\put(.9,2.1){\oval(0.2,0.6)[b]}
\put(.8,2.1){\vector(0,1){5}}
\put(4,3.5){\vector(1,0){2.5}}
\put(4.3,4){\parbox[b]{2cm}{\rm Connes}}
\put(7.5,3.5){\parbox{5cm}{\rm gravity \\${}$ +
Yang-Mills-Higgs ansatz\\${}$ + constraints}}
\put(-0.9,5.5){\parbox{2cm}{\rm Connes}}

\put(0,7.8){\parbox{2cm}{\rm noncommutative
geometry}}
\put(4,7.8){\vector(1,0){2.5}}
\put(8,7.8){\parbox{5cm}{??}}

\end{picture}
\end{center}
\caption{Deriving the
Yang-Mills-Higgs ansatz from gravity}
\end{figure}
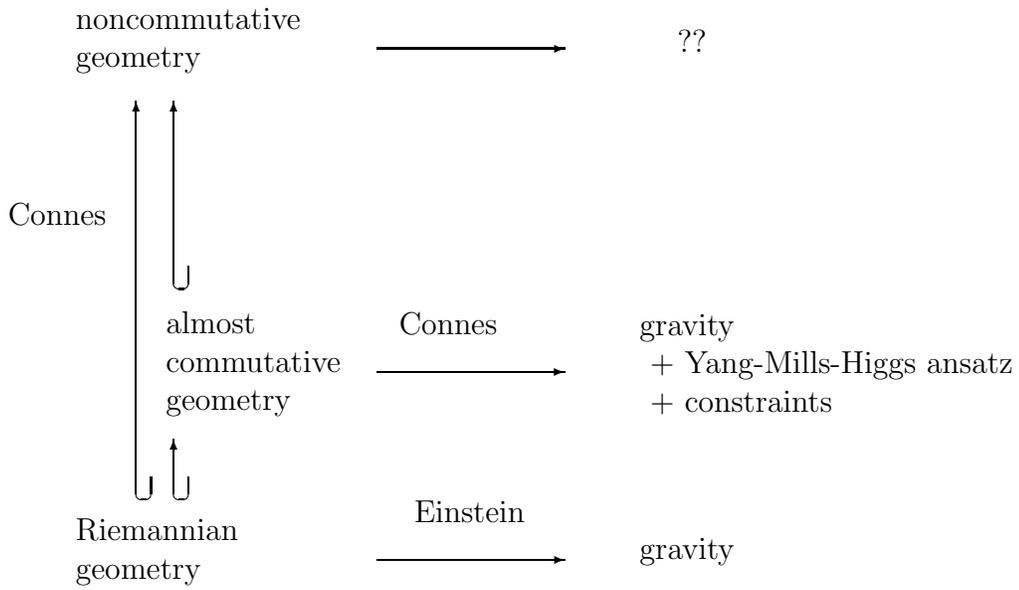

\begin{figure}[h]
\epsfxsize=11cm
\hspace{2.1cm}
\epsfbox{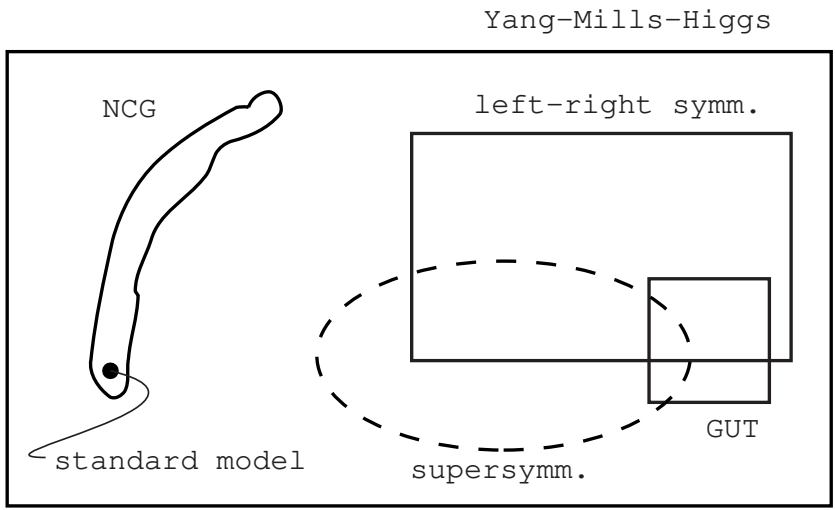}
\caption{Constraints inside the ansatz}
\end{figure}

\begin{figure}[h]
\epsfxsize=11cm
\hspace{2.2cm}
\epsfbox{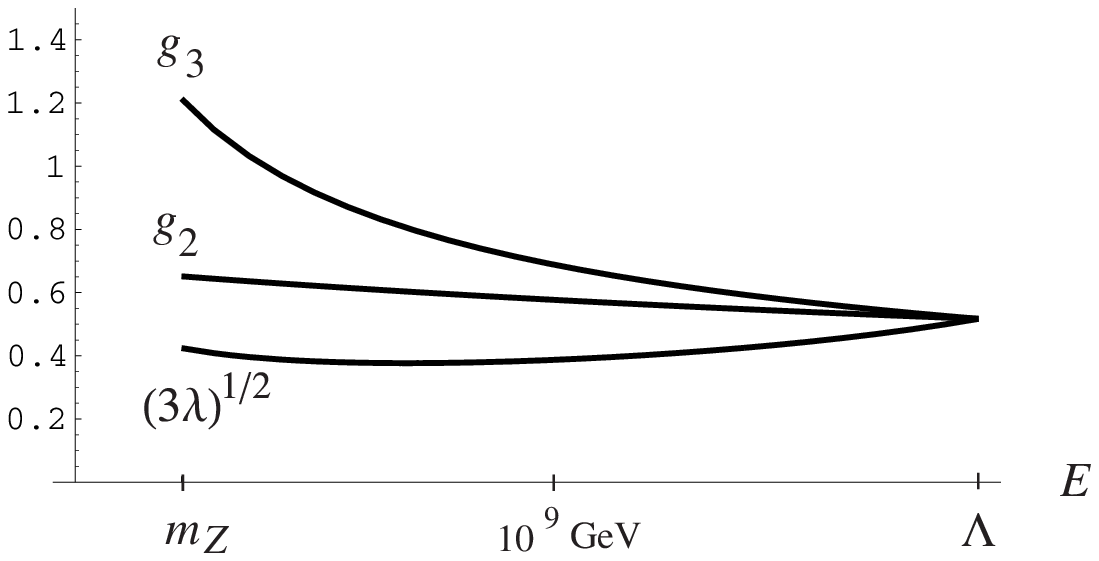}
\caption{Running coupling constants}
\end{figure}


\begin{thebibliography}{47}

\bibitem{data}
The Particle Data Group, {\it Particle Physics Booklet}
and { http://pdg.lbl.gov}
\bibitem{connes}
A. Connes, {\it Noncommutative Geometry}, Academic
Press (1994)\\
A. Connes,
{\it Noncommutative geometry and reality},
J. Math. Phys. 36 (1995) 6194
\bibitem{grav}
A. Connes, {\it Gravity coupled with matter and the
foundation of noncommutative geometry},
hep-th/9603053, Comm. Math. Phys. 155 (1996) 109
\bibitem{cc}
A. Chamseddine \& A. Connes, {\it The
spectral action principle}, hep-th/9606001,
Comm. Math. Phys.186 (1997) 731
\bibitem{john}
J. Madore, {\it An Introduction to Noncommutative
Differential Geometry and Its Physical Applications},
Cambridge University Press (1995)
\bibitem{further}
Further reading:\\
L. O'Raifeartaigh, {\it Group Structure of
Gauge Theories}, Cambridge University Press (1986)\\
C. P. Mart\'\i n, J. M. Gracia-Bond\'\i a \& J. C. V\'arilly,
{\it The standard model as a noncommutative geometry:
the low mass regime},
hep-th/9605001, Phys. Rep. 294 (1998) 363 \\
G. Landi, {\it An Introduction to Noncommutative
Spaces and Their Geometry}, hep-th/9701078,
Springer (1997)\\
J. M. Gracia-Bond\'\i a, J. C. V\'arilly \& H. Figueroa,
{\it Elements of Noncommutative Geometry},
Birkh\"auser (2000)\\
D. Kastler, {\it Noncommutative geometry and fundamental physical
interactions: the Lagrangian level},  J. Math. Phys. 41 (2000) 3867\\
F. Scheck, W. Werner \& H. Upmeier (eds.), {\it
Noncommutative Geometry and the Standard Model of Elementary
Particle Physics}, Lecture notes in physics 596, Springer (2002)\\
T. Sch\"ucker, {\it Forces from Connes' geometry}, in `Topology and
Geometry in Physics', eds.: E. Bick \& F. Steffen, hep-th/0111236,
Lecture notes in physics, Springer, to appear
\end{thebibliography}
\end{document}